\documentclass[a4paper,11pt]{article}
\usepackage{pos}

\title{Mono and stereo performance of the two SST-1M telescope prototypes}
 \ShortTitle{SST-1M performance}

\author*[a]{J.~Jury\v{s}ek}
\author[a]{T.~Tavernier}
\author[a, b]{V.~Novotn\'{y}}
\author[c]{M.~Heller}
\author[a]{D.~Mandat}
\author[a]{M.~Pech}
\author[c]{C.~Alispach}
\author[d,e]{A.~Araudo}
\author[f]{V.~Beshley}
\author[a]{J.~Blazek}
\author[g]{J.~Borkowski}
\author[h]{S.~Boula}
\author[i]{T.~Bulik}
\author[c]{F.~Cadoux}
\author[h]{S.~Casanova}
\author[a]{A.~Christov}
\author[j]{L.~Chytka}
\author[c]{D.~della Volpe}
\author[c]{Y.~Favre}
\author[k]{L.~Gibaud}
\author[h]{T.~Gieras}
\author[j]{P.~Hamal}
\author[j]{M.~Hrabovsky}
\author[l]{M.~Jel\'inek}
\author[d]{V.~Karas}
\author[m]{E.~Lyard}
\author[h]{E.~Mach}
\author[h]{W.~Marek}
\author[j]{S.~Michal}
\author[h]{J.~Micha{\l}owski}
\author[g]{R.~Moderski}
\author[c]{T.~Montaruli}
\author[g]{A.~Muraczewski}
\author[a]{S.~R.~Muthyala}
\author[c]{A.~Nagai}
\author[h]{K.~Nalewajski}
\author[n]{D.~Neise}
\author[h]{J.~Niemiec}
\author[k]{M.~Niko{\l}ajuk}
\author[o]{M.~Ostrowski}
\author[a]{M.~Palatka}
\author[a]{M.~Prouza}
\author[p]{P.~Rajda}
\author[a]{P.~Schovanek}
\author[q]{K.~Seweryn}
\author[m]{V.~Sliusar}
\author[o]{{\L}.~Stawarz}
\author[h]{J.~\'{S}wierblewski}
\author[h]{P.~\'{S}wierk}
\author[l]{J.~\v{S}trobl}
\author[a]{J.~V\'icha}
\author[m]{R.~Walter}
\author[o]{A.~Zagda\'{n}ski}
\author[o]{K.~Zi{\c e}tara}

\scriptsize
\affiliation[]{
$^a$FZU - Institute of Physics of the Czech Academy of Sciences, Na Slovance 1999/2, Prague 8, Czech Republic,
$^b$Charles University, Faculty of Mathematics and Physics, Institute of Particle and Nuclear Physics, Prague, Czech Republic,
$^c$DPNC - Universit\'e de Gen\`eve, 24 Quai Ernest Ansermet, CH-1211 Gen\`eve,  Switzerland,
$^d$Astronomical Institute of the Czech Academy of Sciences, Bo\v{c}n\'i II 1401, CZ-14100 Prague, Czech Republic,
$^e$ELI Beamlines, Institute of Physics, Czech Academy of Sciences, CZ-25241 Doln\'i B\v{r}e\v{z}any, Czech Republic,
$^f$Pidstryhach Institute for Applied Problems of Mechanics and Mathematics, National Academy of Sciences of Ukraine, 3-b Naukova St., 79060, Lviv, Ukraine,
$^g$Nicolaus Copernicus Astronomical Center, Polish Academy of Sciences,  ul. Bartycka 18, 00-716 Warsaw, Poland,
$^h$Institute of Nuclear Physics Polish Academy of Sciences, PL-31342 Krakow, Poland,
$^i$Astronomical Observatory, University of Warsaw, Al. Ujazdowskie 4, 00-478 Warsaw, Poland,
$^j$Palacky University Olomouc, Faculty of Science, 17. listopadu 50, Olomouc, Czech Republic,
$^k$Faculty of Physics, University of Bia{\l}ystok, ul. K. Cio{\l}kowskiego 1L, 15-245 Bia{\l}ystok, Poland,
$^l$Astronomical Institute of the Czech Academy of Sciences, Fri\v{c}ova 298, CZ-25165 Ond\v{r}ejov, Czech Republic,
$^m$D\'epartement d'Astronomie, Universit\'e de Gen\`eve, Chemin d'Ecogia 16, CH-1290 Versoix, Switzerland,
$^n$ETH Zurich, Institute for Particle Physics and Astrophysics, Otto-Stern-Weg 5, 8093 Zurich, Switzerland,
$^o$Astronomical Observatory, Jagiellonian University, ul. Orla 171, 30-244 Krakow, Poland,
$^p$AGH University of Science and Technology, al.Mickiewicza 30,  30-059 Krakow, Poland,
$^q$Centrum Bada{\'n} Kosmicznych Polskiej Akademii Nauk,  18a Bartycka str., 00-716 Warsaw, Poland,
}

\emailAdd{jurysek@fzu.cz}

\abstract{
The Single-Mirror Small-Sized Telescope, or SST-1M, was originally developed as a prototype of a small-sized telescope for CTA, designed to form an array for observations of gamma-ray-induced atmospheric showers for energies above 3 TeV. A pair of SST-1M telescopes is currently being commissioned at the Ond\v{r}ejov Observatory in the Czech Republic, and the telescope capabilities for mono and stereo observations are being tested in better astronomical conditions. The final location for the telescopes will be decided based on these tests. In this contribution, we present a data analysis pipeline called \texttt{sst1mpipe}, and the performance of the telescopes when working independently and in a stereo regime.
}

\ConferenceLogo{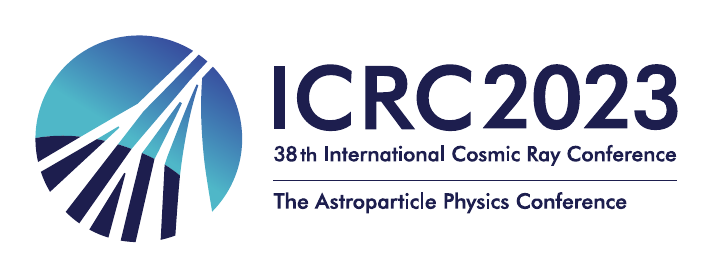}

\FullConference{%
38th International Cosmic Ray Conference (ICRC2023)\\
  26 July - 3 August, 2023\\
  Nagoya, Japan}

\begin{document}
\maketitle

\section{Introduction}

The Single-Mirror Small-Sized-Telescope (SST-1M) was originally designed for observation of gamma rays of energies above 3 TeV as an integral part of the future next-generation gamma-ray observatory - Cherenkov Telescope Array (CTA). The layout of the telescope's 4-m multi-segment mirror follows the Davies-Cotton concept. The telescope camera features a fully digital trigger and readout system. It is based on the SiPM technology, which allows for the telescope operation even under high Night Sky Background (NSB) conditions not suffering from the aging effect like common photo-multipliers \cite{camera_paper, cta_sst1m}.

After the first part of the commissioning phase in Krakow, and despite having been judged fully compliant with CTA requirements, SST-1M was not selected as the final design of small telescopes for CTA. The telescope prototypes were transported in 2021 to Ond\v{r}ejov Observatory in the Czech Republic, where the telescopes are being tested until their final location is decided. The telescopes are installed 155 meters apart, which allows for stereoscopic observation of the atmospheric showers. After the construction of the telescopes was finished in 2022, both telescopes observed showers in pseudo-stereo mode, where the coincident events were matched in post-processing using a relatively wide time coincidence window. At the beginning of 2023, the time stamps were precisely synchronized using the White Rabbit network, allowing for the first tests of real-time stereo observations.

The Ond\v{r}ejov Observatory located near Prague serves as an ideal and well-accessible test bench for the telescope operations, allowing for fast interventions if needed. Its relatively low altitude of 510 m a.s.l. challenges, however, the telescope capabilities for observation of sources at lower energies. In this contribution, we present a data analysis pipeline, for both mono and stereoscopic event reconstruction. We present the status of Monte Carlo (MC) - data agreement and discuss telescope performance and prospects for scientific observations at low-altitude sites. The telescope calibration strategy and results of the first observations of the Crab Nebula are presented in \cite{tavernier_sst1m_2023}.

\section{\texttt{sst1mpipe}}

\texttt{sst1mpipe} is a data and MC analysis pipeline for SST-1M, currently in intensive development. It is an independent standalone implementation with a basic structure inspired by \texttt{lstchain} \cite{ruben_lopez_coto_2022_6344674}, and stereo reconstruction inspired by \texttt{magic-cta-pipe}\footnote{https://github.com/cta-observatory/magic-cta-pipe}, from which it diverges significantly in certain SST-1M-specific tasks. For low-level calibration or event handling \texttt{sst1mpipe} relies on some functionalities of \texttt{digicampipe}\footnote{https://github.com/cta-sst-1m/digicampipe/tree/master}. The current version of the pipeline is heavily based on \texttt{ctapipe} \cite{karl_kosack_2021_5720333}, low-level data analysis framework of CTA, and closely follows \texttt{ctapipe} data model. Our goal is to produce the output DL3 data (photon lists + Instrument Response Functions (IRFs)) in the modern format following GADF standards \cite{universe7100374} making it compatible with modern tools for high-level data analysis like \texttt{gammapy} \cite{axel_donath_2021_5721467}. 

\begin{figure}[!t]
\centering
\begin{tabular}{c}
\includegraphics[width=.98\textwidth]{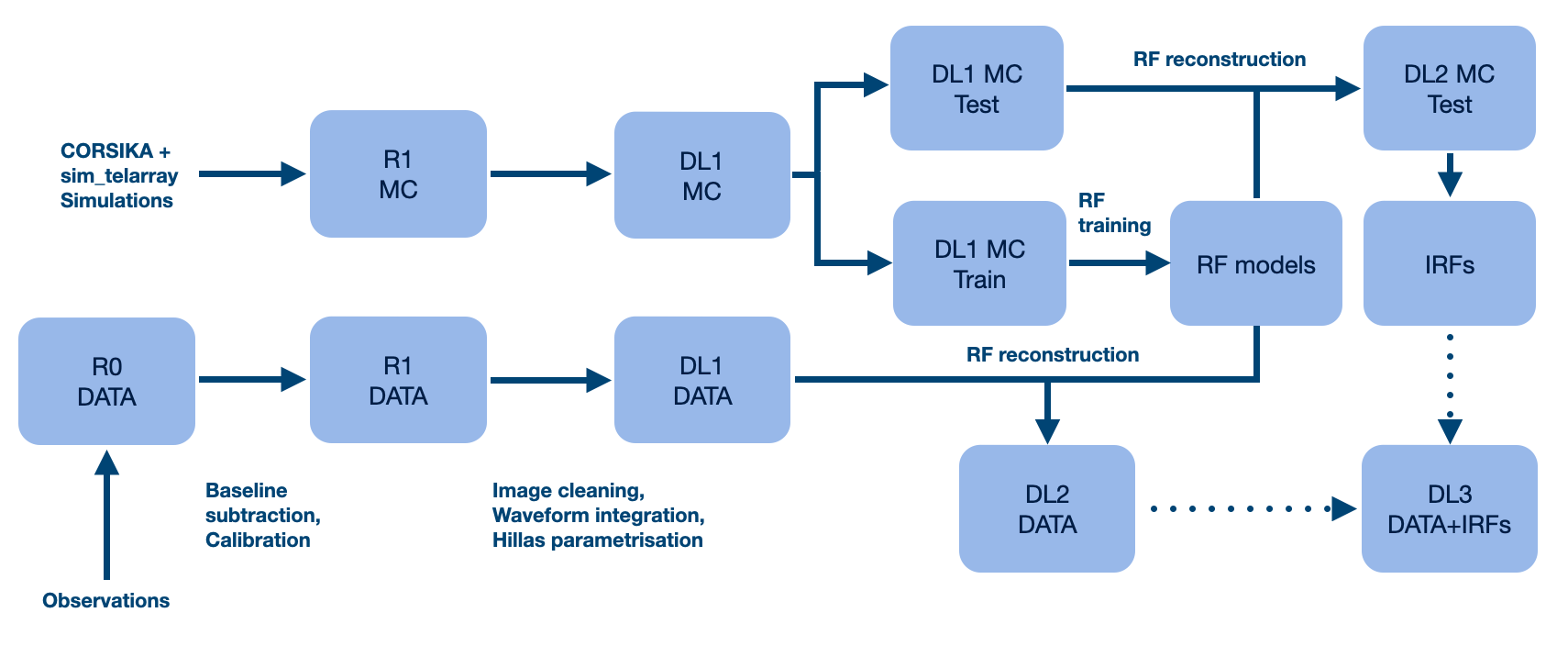}\\
\end{tabular}
\caption{Scheme of the \texttt{sst1mpipe}. Dotted lines represent not-yet-implemented parts of the pipeline.}
\label{fig.sst1mpipe}
\end{figure}

The functionality of the pipeline, schematically shown in Figure~\ref{fig.sst1mpipe}, can be split into two main branches: MC and real data processing, where MC is used to produce IRFs and to train Random Forests (RFs) using \texttt{scikit-learn} framework \cite{scikit-learn} for real data reconstruction. The data collected by the telescopes contain for each triggered event a raw wave-form in all camera pixels (R0 data level), that needs to be calibrated and converted to photo-electrons (R1) \cite{tavernier_sst1m_2023}. MC simulations are produced in \texttt{CORSIKA} \cite{1998cmcc.book.....H} and \texttt{sim\_telarray} \cite{BERNLOHR2008149} software packages, which provide calibrated wave-forms (R1) for each triggered event. To extract the shower image, these are integrated using LocalPeakWindowSum integrator \cite{karl_kosack_2021_5720333}. The tailcut cleaning method is adopted in the pipeline for the removal of the noisy pixels containing only the Night Sky Background in the integrated images. After the cleaning, the pixels containing signal from the Cherenkov photons are parametrized to describe the shape and orientation of the shower image (the so-called Hillas parameters) \cite{1985ICRC....3..445H}. DL1 data level contains shower images, cleaning masks, and Hillas parameters.

DL1 MC events are then separated into training and testing data sets. The training data set is used for training a RF classifier for gamma/hadron separation and an RF regressor for energy reconstruction. For direction reconstruction in mono analysis, we adopt the so-called \textit{disp} method \cite{LESSARD20011} assuming that the source lies on the main axis of the shower image. Then an RF regressor is trained to reconstruct the distance of the source from the image centroid (\textit{disp norm}), subsequently followed by an RF classifier to determine on which side of the centroid the source lies (\textit{disp sign}). DL2 data contain reconstructed energies, arrival directions, and \textit{gammaness}, which is a parameter with a numeric value between 0 and 1 representing how much is a given shower image gamma-like.

In stereo reconstruction, where the shower image from each telescope is available for each event, we first perform preliminary reconstruction of the height of the shower maximum and impact parameter for each telescope using the axis crossing method \cite{Hofmann:1999rh}. These parameters are then used as additional features for RF training. Energy regressor and gamma/hadron classifier are trained per telescope, and their outputs in the reconstruction of the testing MC sample and the real data are averaged using the image intensity as weight. For the stereo-direction reconstruction, the pipeline follows the MARS-like approach \cite{ALEKSIC201676}. First, the distance of the source from the image centroid is reconstructed for each telescope. Then both head-tail orientations of the shower are considered for each telescope, and four tentative source coordinates are calculated. In the last step, the combination providing the closest angular distance is found and the final reconstructed position is calculated as a weighted average of the source coordinates from both telescopes.


\section{Monte Carlo simulations of SST-1M}

A detailed MC model of SST-1M-1 (including the reflectivity of the mirror, the transmissivity of the camera window, the shadowing of the telescope structure, and camera electronics) was determined and validated on data during the first commissioning phase while the telescope was installed at IFJ Krakow \cite{sst1m_mc_2019, 2020JInst..15P1010A}. The second telescope SST-1M-2, however, differs from SST-1M-1 in several aspects (e.g. different transmissivity of the camera entrance window \cite{DeAngelis:2683289}, or different gain drop \cite{Nagai_2019}), and we thus repeated the study described in \cite{sst1m_mc_2019}, this time including also detailed modeling of the optical Point Spread Function (PSF) for both telescopes. 

\begin{figure}[!t]
\centering
\begin{tabular}{ccc}
\includegraphics[width=.35\textwidth]{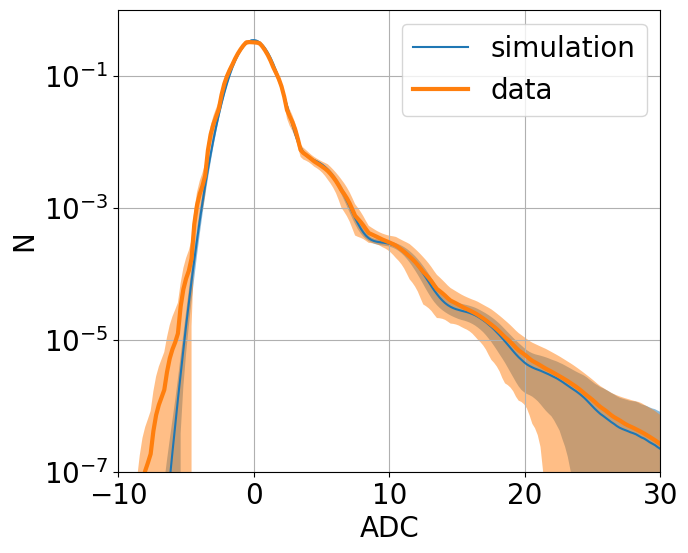} & \includegraphics[width=.31\textwidth]{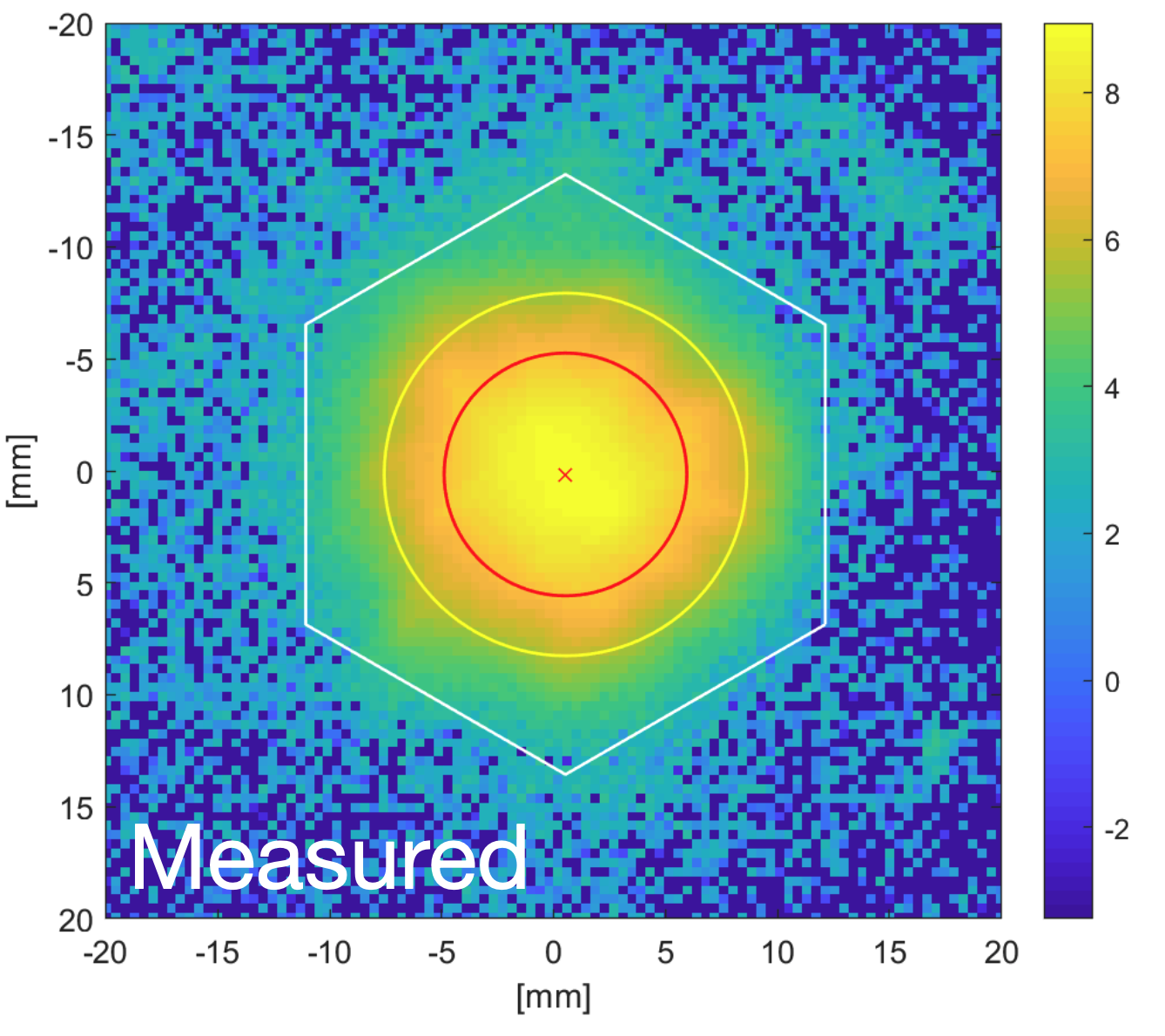} & \includegraphics[width=.29\textwidth]{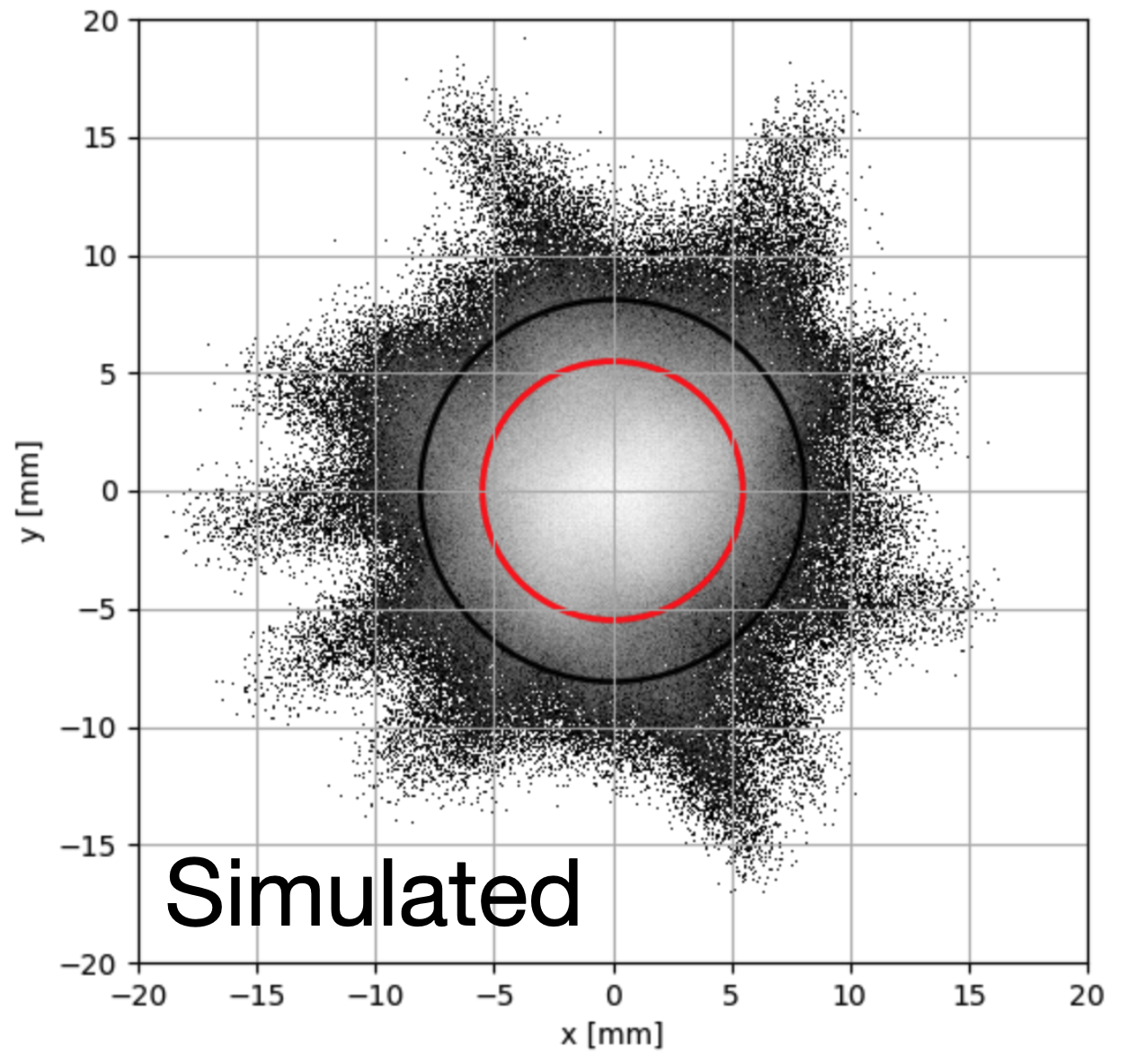} \\
\end{tabular}
\caption{\textit{Left:} Distributions of raw all-pixel averaged ADC counts for dark run data and simulations. \textit{Middle:} Measured on-axis optical PSF of SST-1M-2. The red circle marks D80, yellow circle marks D100, set by definition as 16.2 mm. The white hexagon shows the size of a single pixel. \textit{Right:} Simulated optical PSF for SST-1M-2.}
\label{fig.raw_adc}
\end{figure}

We first simulated dark events for different values of gain, noise, and dark count rate, comparing the distributions of ADC counts (all-pixel average) in the raw waveforms with dark data taken with the camera lid closed. Left Figure~\ref{fig.raw_adc} shows MC-data comparison for SST-1M-2 for the best matching parameters. We note that the negative non-gaussian tail of the distribution is still under investigation. Middle Fig.~\ref{fig.raw_adc} shows the on-axis optical PSF measured for SST-1M-2. We simulated a point-like light source at infinity in \texttt{sim\_telarray}, assuming radial symmetry of the PSF. The measured on-axis spot diameter D80 (80\% of the detected energy) is 9.6~mm and 10.8 mm for SST-1M-1 and SST-1M-2, respectively.

Using the detailed MC model, we prepared a large-scale MC production for both telescopes including diffuse gammas and diffuse protons used for RF training, and point-like gammas for testing and performance evaluation. We simulated both telescopes at once, keeping a minimum trigger multiplicity equal to one, which allowed us to estimate the performance of both telescopes working in a mono regime and in stereo mode. Vertical Aerosol Optical Depth in the atmospheric profile used in the \texttt{sim\_telarray} simulations was set to 0.2, which is a typical value for the Ond\v{r}ejov Observatory based on our preliminary atmospheric study. NSB level was estimated by comparison of simulated and measured rate scans (see \cite{sst1m_mc_2019} for details), and fixed on the level of 100~Hz per pixel, representing a typical dark NSB rate at the site measured with the mono trigger. Studies of the SST-1M performance under higher NSB conditions are currently ongoing. We note that the NSB-induced trigger rate is expected to be lower when the stereo trigger is used, which would allow for lowering the trigger threshold and therefore also the energy threshold in the analysis. The simulations were performed for zenith angles of $20^\circ$, $40^\circ$, and $60^\circ$, in order to understand the zenith angle dependency of the energy threshold and sensitivity of the instrument. 

\begin{figure}[!t]
\centering
\begin{tabular}{ccc}
\includegraphics[width=.33\textwidth]{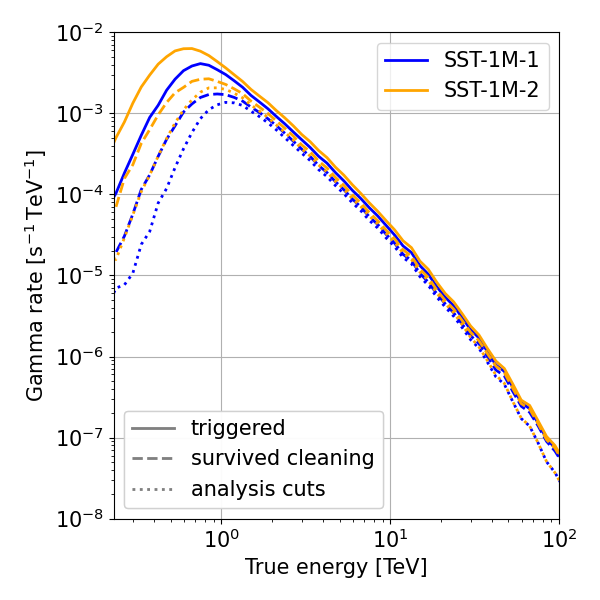} & \includegraphics[width=.33\textwidth]{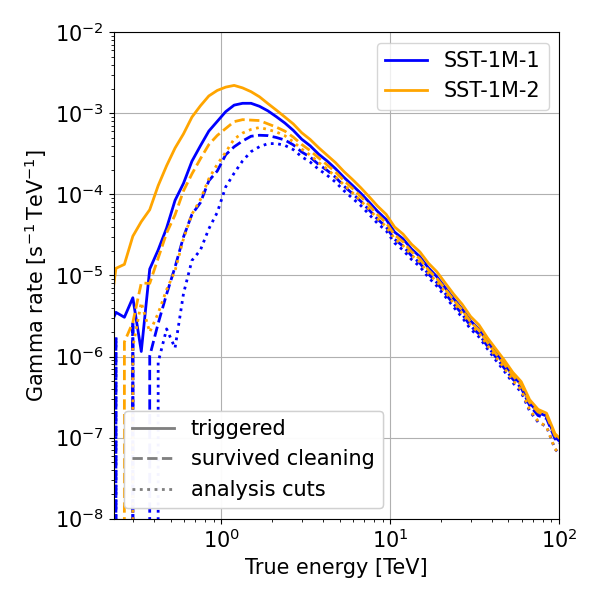} & \includegraphics[width=.33\textwidth]{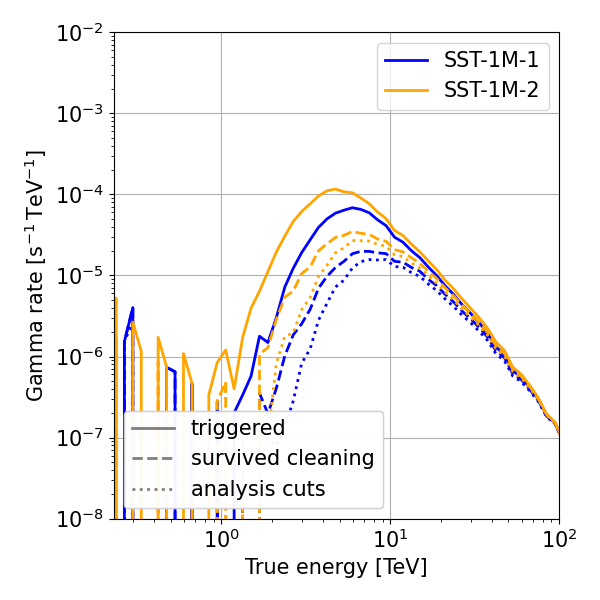} \\
\end{tabular}
\caption{Differential rate of point-like gamma rays with Crab Nebula spectrum seen with both telescopes (in mono regime) for $20^\circ$ (\textit{left}), $40^\circ$ (\textit{middle}), and $60^\circ$ (\textit{right}) zenith angle. Event rates for different stages of the analysis are shown: All triggered events (solid line), all events that survived cleaning (dashed line), and events that survived typical quality cuts used in the analysis (dotted line).}
\label{fig.diff_event_rate}
\end{figure}

The energy threshold is usually defined as the MC energy of primary gamma rays (called hereafter "True energy"), for which the differential event rate of gamma rays with a given spectral energy distribution reaches its maximum. Figure~\ref{fig.diff_event_rate} shows the rate of point-like gamma rays weighted on the Crab Nebula spectrum \cite{ALEKSIC201530} for all three simulated zenith angles and different stages of the analysis. The energy threshold strongly depends on the zenith angle, starting at about 1~TeV for $20^\circ$ up to about 5~TeV for $60^\circ$ at the trigger level.

\begin{figure}[!t]
\centering
\begin{tabular}{c}
\includegraphics[width=.99\textwidth]{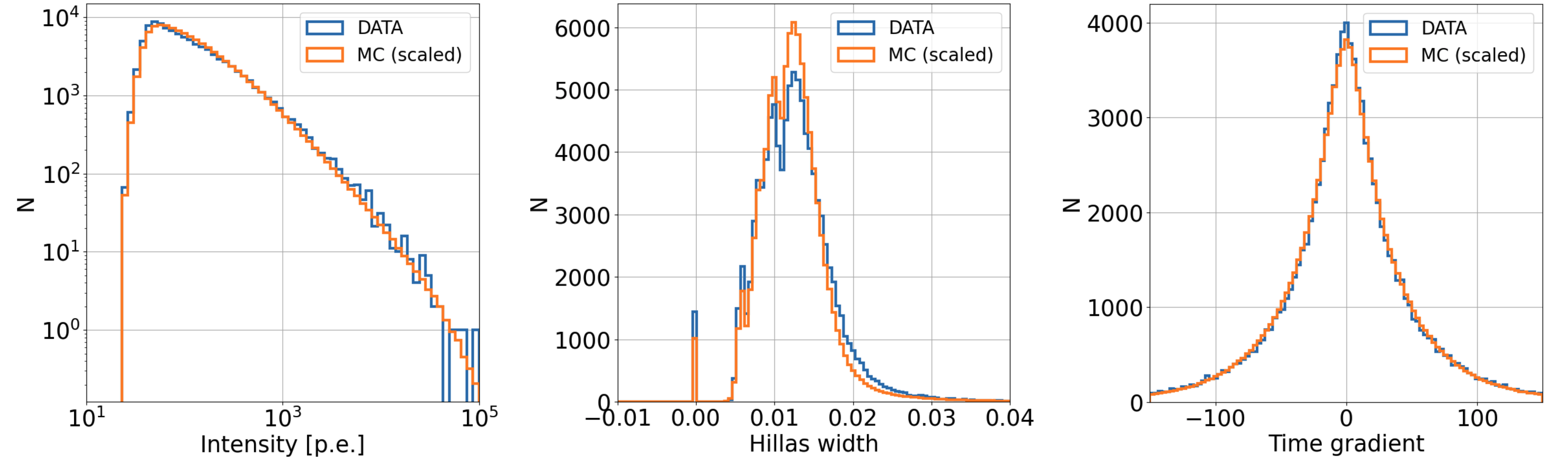}\\
\end{tabular}
\caption{Comparison of selected Hillas parameters for data taken on June 12, 2023 at zenith angles between $18^\circ$ and $22^\circ$ with diffuse proton MC re-weighted on the CR spectrum. The distributions of MC simulated events were scaled by a factor of 1.04 to account for the actual atmospheric transparency.}
\label{fig.mc_data_hillas}
\end{figure}

Figure~\ref{fig.mc_data_hillas} demonstrates a preliminary MC/data comparison of Hillas parameter distributions 
for particular run taken on June 12, 2023 at zenith angles between $18^\circ$ and $22^\circ$, and diffuse proton MC simulated at $20^\circ$ zenith angle. The spectrum of simulated protons was re-weighted on DAMPE p+He spectrum \cite{DeMitri:2021jog} to correct for the Cosmic-Ray (CR) composition. We note that the distributions are of all events that survived image cleaning and no quality cuts were applied. The data was taken in dark conditions and MC had to be scaled by only a factor of $1.04$ to account for the difference between simulated atmospheric transparency (so far fixed) and the real one at the time of the observation. The MC-data agreement is good enough to run the reconstruction of real events with RFs trained on MC (for more details and preliminary results see \cite{tavernier_sst1m_2023}), but there is still room for improvement, especially in simulated shower width, which is one of the most important features for the RF classifier. Unlike simulations, real shower images are affected by a presence of stars in the FoV, locally increasing the level of NSB, which can propagate into fake islands of pixels left in the image after cleaning, especially in the case of fixed cleaning tailcuts currently adopted in the analysis. Inhomogeneous or variable NSB due to light pollution from nearby villages or Prague may have a similar effect. An alternative cleaning method with adaptive tailcuts reflecting the local level of NSB and exploiting shower timing information in the telescope data is currently in development.

\section{Performance of SST-1M at low altitude}

\begin{figure}[!t]
\centering
\begin{tabular}{cc}
\includegraphics[width=.47\textwidth]{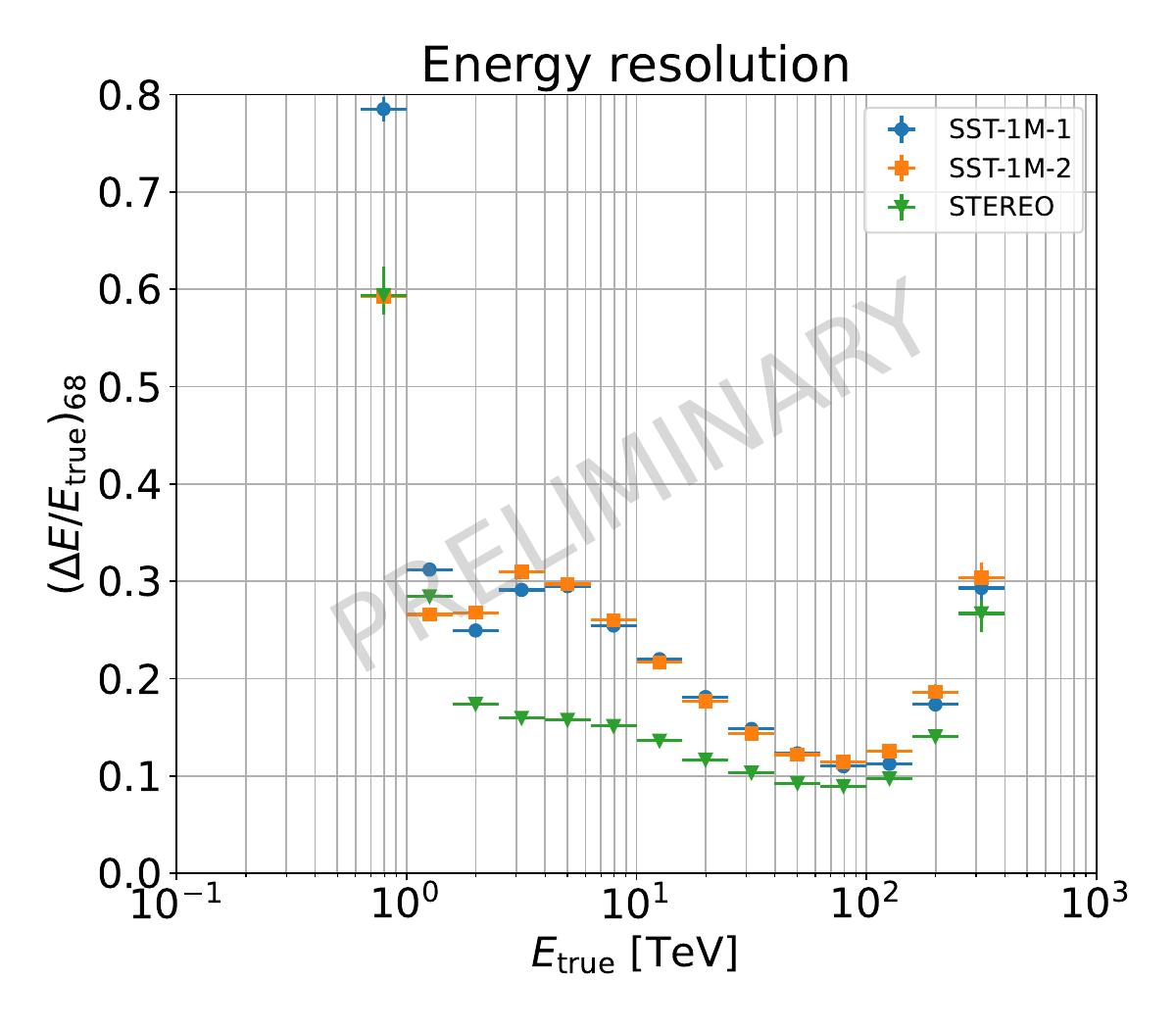} & \includegraphics[width=.47\textwidth]{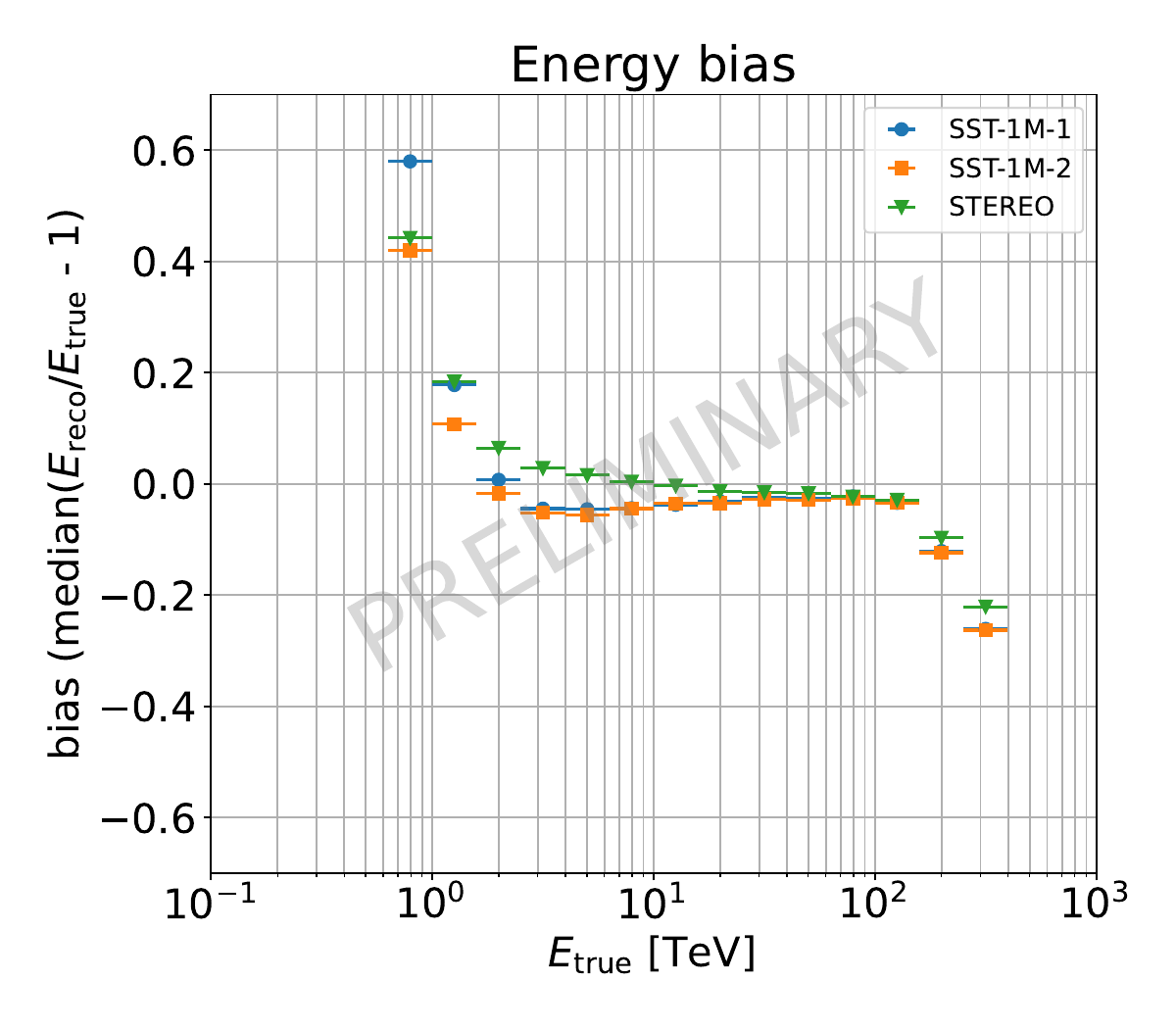} \\
\includegraphics[width=.47\textwidth]{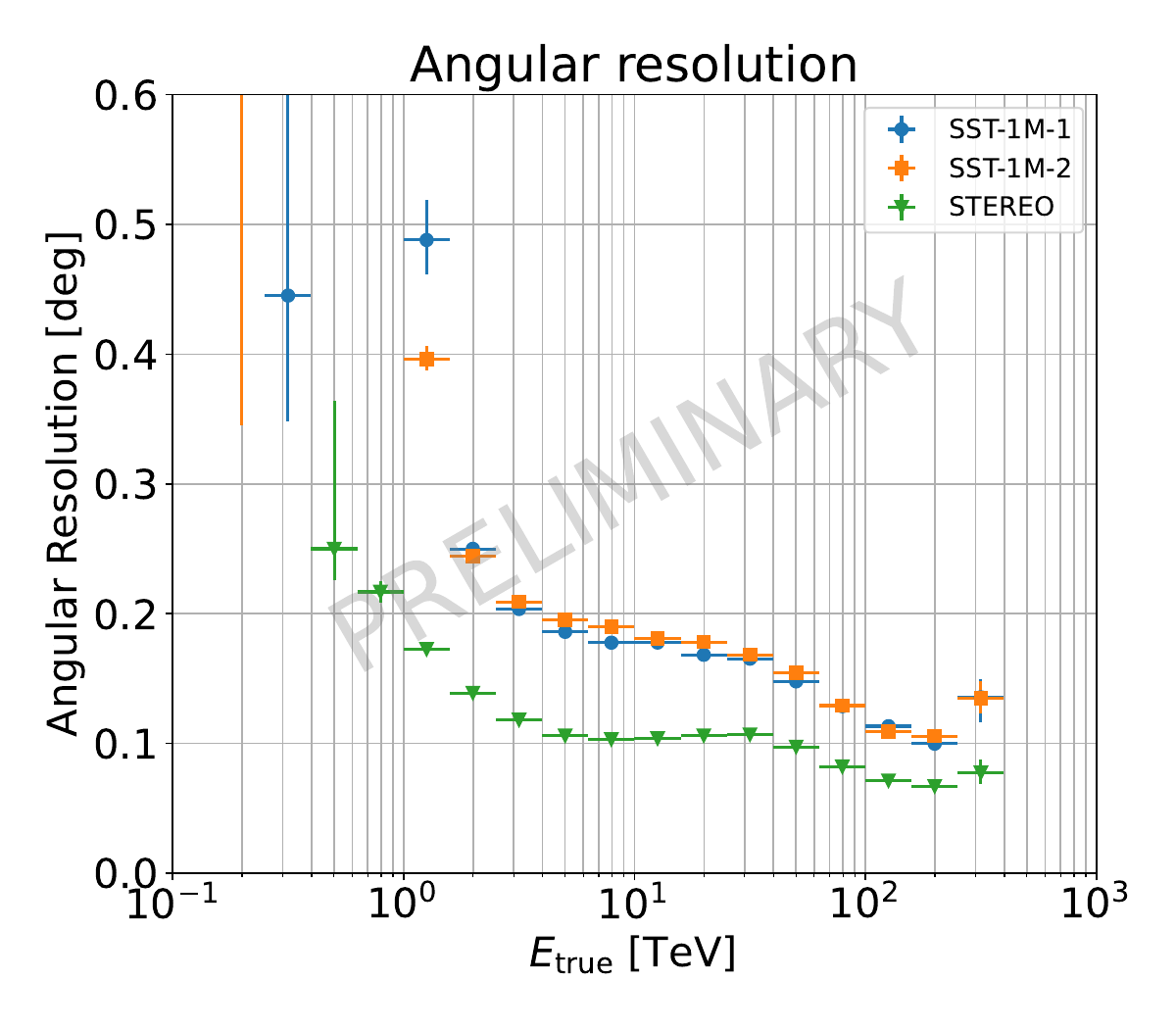} & \includegraphics[width=.47\textwidth]{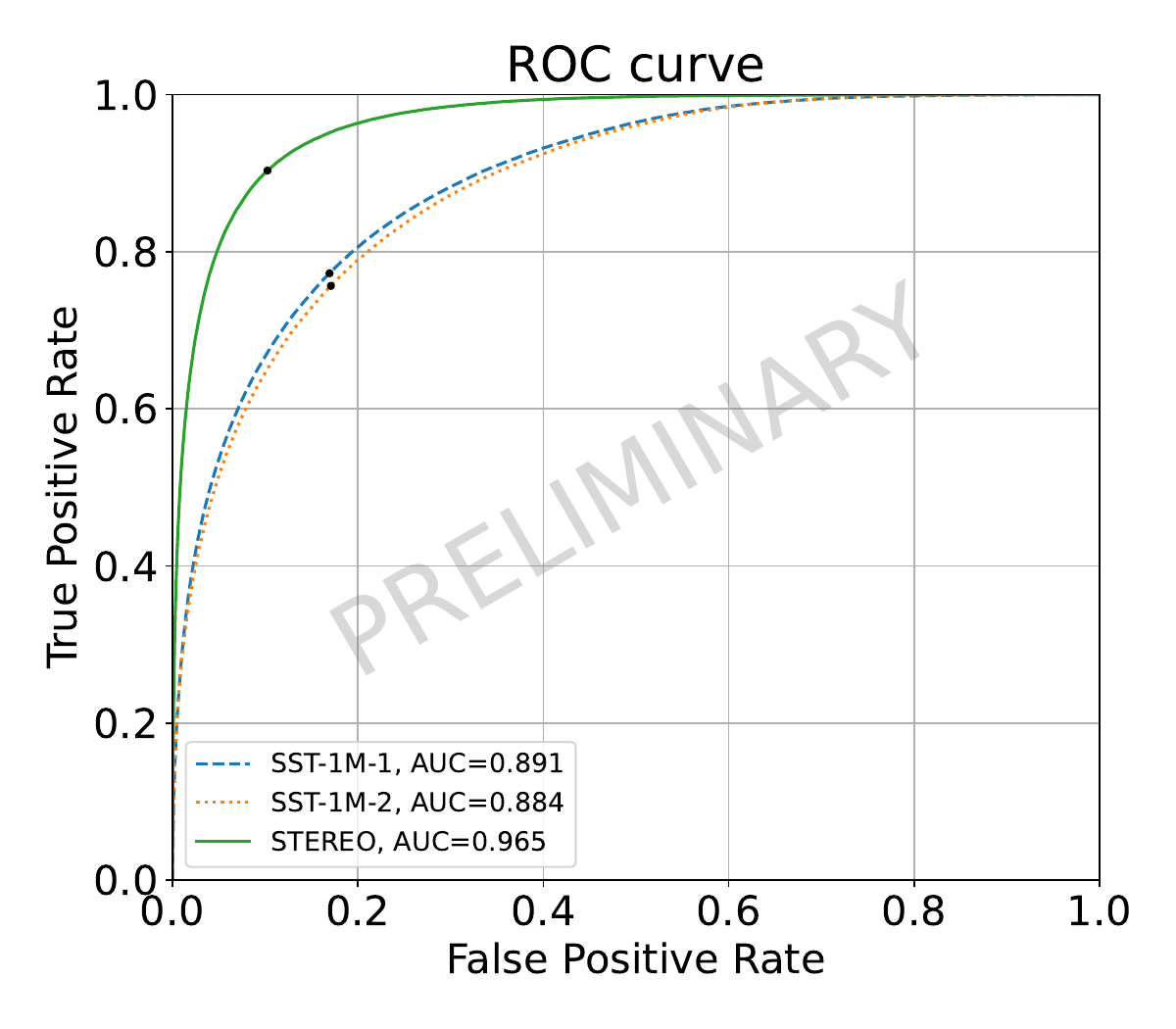} \\
\end{tabular}
\caption{Mono and stereo performance of SST-1M telescopes at the Ond\v{r}ejov Observatory for $20^\circ$ zenith angle. \textit{Top left:} energy resolution, \textit{Top right:} energy bias, \textit{Bottom left:} angular resolution, \textit{Bottom right:} Receiver Operation Characteristics (ROC) quantifying the power of gamma/hadron separation. Black dots on the ROC curves mark \textit{gammaness} $>0.6$ cut, which is applied to the DL2 testing data set in this study.}
\label{fig.performance_20_deg}
\end{figure}

Figure~\ref{fig.performance_20_deg} shows the preliminary performance at $20^\circ$ zenith angle for both SST-1Ms working individually in mono and in the stereo mode at the altitude of 510~m.a.s.l. at the geographical location of the Ond\v{r}ejov Observatory. In this analysis, we applied relatively weak quality cuts on the reconstructed DL2 point-like gamma MC data set: \textit{intensity} > 50 p.e., \textit{leakage2} < 0.2, and \textit{gammaness} > 0.6, where \textit{leakage2} represents a fraction of shower pixels in the two outermost rings of the camera pixels. The energy resolution improves with energy until $\approx100$ TeV, reaching $\approx10\%$. The reconstruction of events with higher energies starts to be limited by the containment of the shower images within the camera FoV. Stereo reconstruction improves the performance significantly at energies below 10~TeV, where the impact parameter is not well reconstructed in the mono analysis. The energy bias is below $5\%$ from energies slightly above the energy threshold up to about 200~TeV. The angular resolution in mono is below $\approx0.2^\circ$ above the energy threshold for both telescopes, suffering mostly from wrongly reconstructed \textit{disp sign} for low energy showers. This issue is removed in stereo reconstruction, where the angular resolution reaches  $\approx0.1^\circ$. 

\begin{figure}[!t]
\centering
\begin{tabular}{cc}
\includegraphics[width=.47\textwidth]{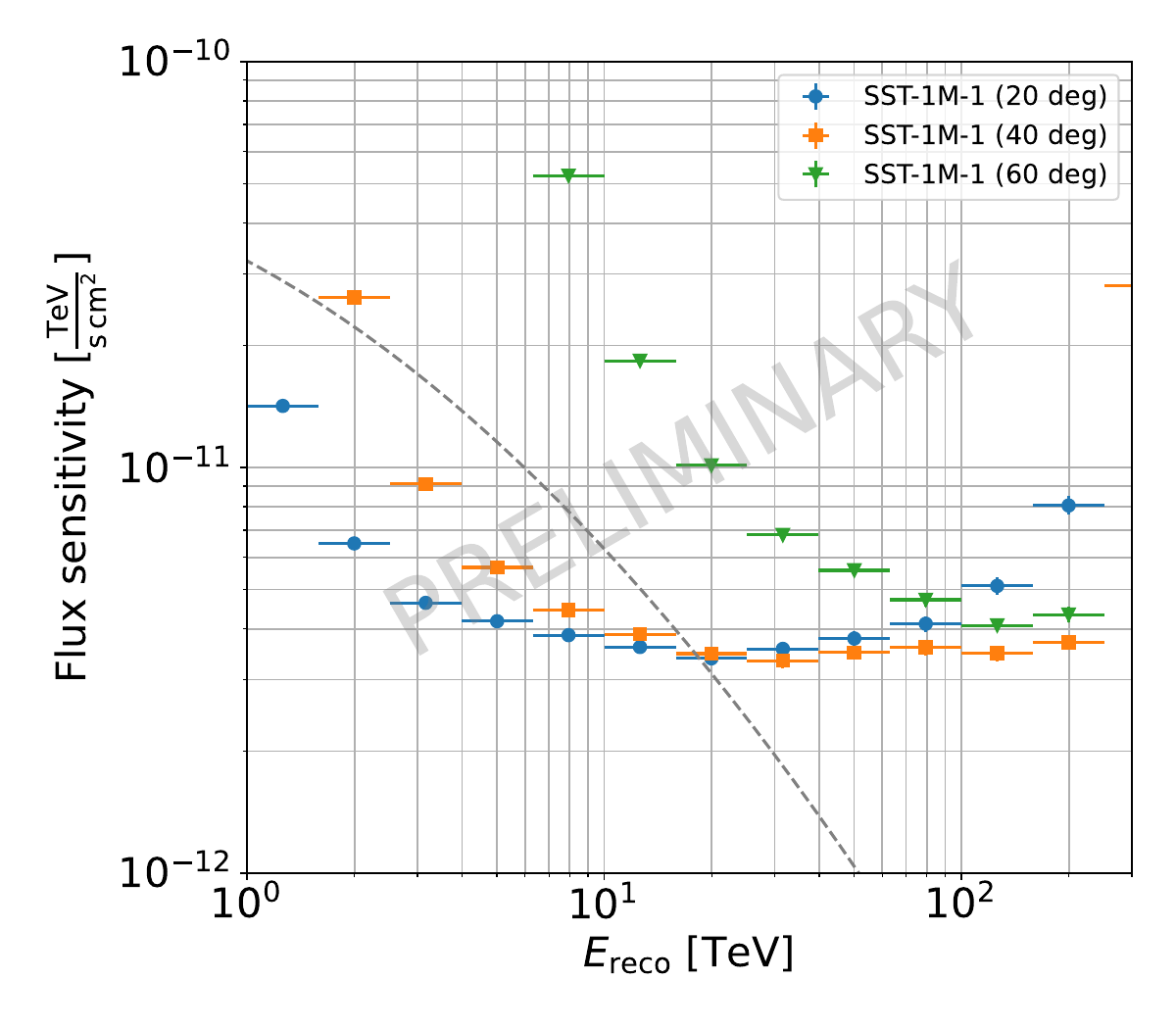} & \includegraphics[width=.47\textwidth]{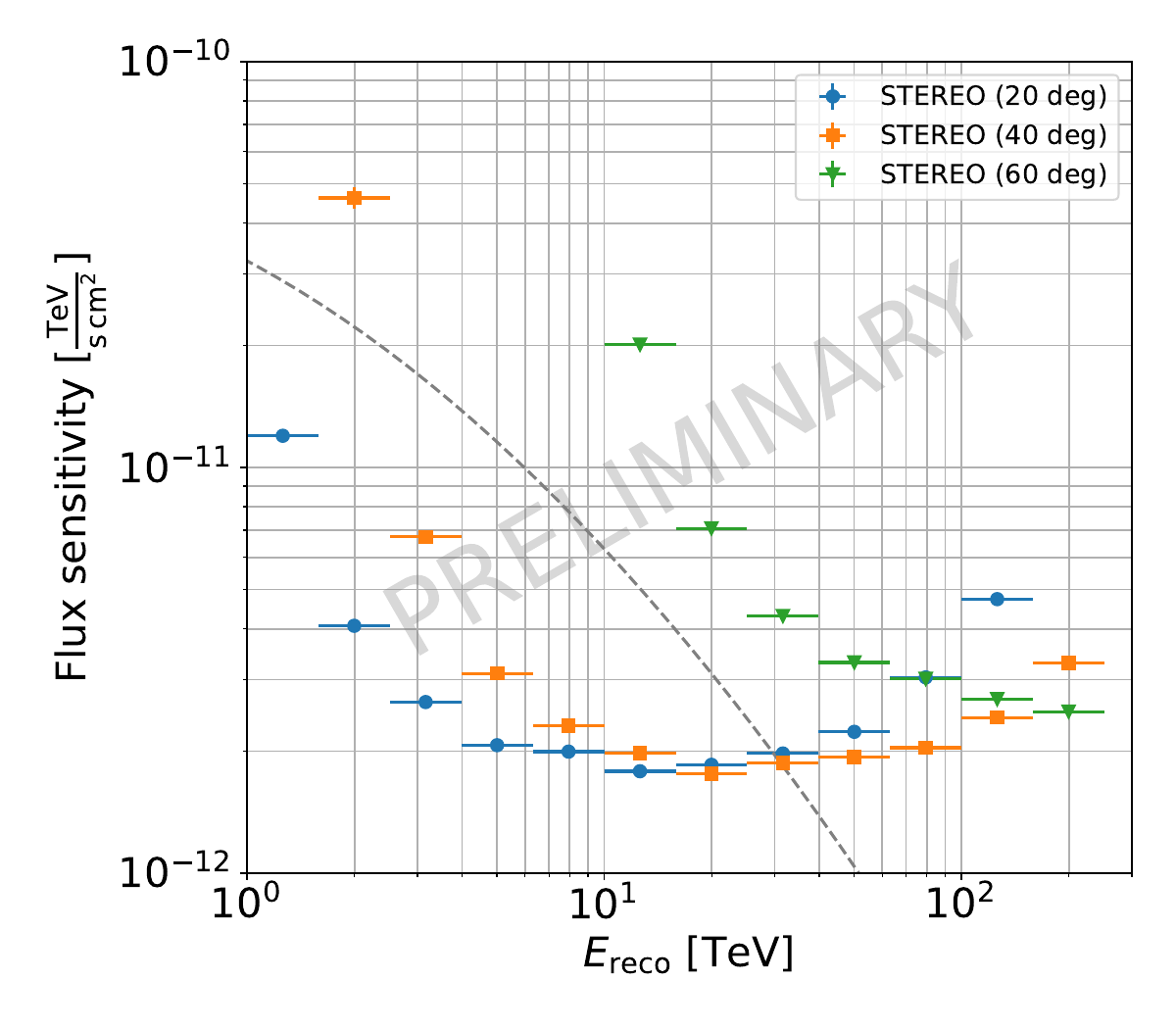} \\
\end{tabular}
\caption{Differential sensitivity of SST-1M-1 telescope in mono (\textit{left}) and for both telescopes in stereo mode (\textit{right}) for 50 hours of observation at different zenith angles. Dashed line marks Crab Nebula spectrum \cite{ALEKSIC201530}.}
\label{fig.sensitivity_zenith}
\end{figure}

The differential sensitivity of SST-1M mono/stereo for different zenith angles and 50 hours of observation is shown in Figure~\ref{fig.sensitivity_zenith}. Besides the same quality cuts discussed above, in this case, we applied energy-dependent \textit{gammaness} cut, which was optimized in each energy bin so that it provides the best detection significance of a point-like source with the Crab Nebula spectrum on the background of diffuse MC protons re-weighted on the CR p+He spectrum. To select the region of interest, we applied a global $\theta^2$ cut equal to the angular resolution averaged over all energy bins. Sensitivity follows the CTA standards, requesting in each energy bin to reach $5\sigma$ detection significance, while the number of excess events in the signal region $N_\mathrm{ex} \geq 10$, and $N_\mathrm{ex}$ $\geq 5\%$ of the number of background events. The stereo observation improves the sensitivity by a factor of $\approx2$. We note, however, that the stereo sensitivity does not take into account the expected performance improvement at low energies due to the NSB trigger rate suppression discussed above. As expected, sensitivity to low energy events becomes worse with the zenith angle due to increased airmass. On the other hand, sensitivity for Ultra-High-Energy gamma rays improves with the zenith angle due to the effect of the Cherenkov cone geometry, which allows for observations of bright galactic PeVatron candidates. We estimated the time needed to reach $5\sigma$ detection of the Crab Nebula at $20^\circ$ zenith angle as $2.0^{+0.2}_{-0.2}$ h and $1.0^{+0.1}_{-0.1}$ h for mono and stereo observation, respectively. At $40^\circ$ zenith angle, the Crab Nebula is expected to be detected in $4.8^{+0.4}_{-0.4}$ h in mono and $2.0^{+0.2}_{-0.2}$ h in stereo. At $60^\circ$ zenith angle, the sensitivity reaches its limitation imposed by the low altitude of the site and the Crab Nebula cannot be detected even within 50 h of observation.

\section{Conclusion}

In this contribution, we presented the current status of the SST-1M project from the perspective of expected physics performance. The analysis pipeline \texttt{sst1mpipe} is now fully capable of performing both MC and data event reconstruction and analysis. Having a precise MC model of both telescopes, we estimated their mono and stereo performance at the Ond\v{r}ejov observatory, showing that despite the low altitude of the site, the telescopes are fully capable of detecting astrophysical gamma-ray sources.

\section*{Acknowledgements}\scriptsize
\vspace{-2mm}
The contribution of the Czech authors is supported by research infrastructure CTA-CZ, LM2023047 MEYS, and Czech Science Foundation, GACR 23-05827S. Funding by the Polish Ministry of Education and Science under project DIR/WK/2017/2022/12-3 is gratefully acknowledged. The contribution of the Départment de Physique Nucléaire et Corpusculare, Faculty de Sciences of the University of Geneva,  1205 Genève is supported by The Foundation Ernest Boninchi,1246 Corsier-CH, and the Swiss National Foundation (166913, 154221, 150779, 143830).
\vspace{-3mm}

\bibliographystyle{JHEPe}
{\footnotesize
\bibliography{references}}

\providecommand{\href}[2]{#2}\begingroup\raggedright\begin{thebibliography}{10}

\bibitem{camera_paper}
M.~Heller et~al.,
  \href{https://doi.org/10.1140/epjc/s10052-017-4609-z}{\emph{Eur. Phys. J. C}
  {\bfseries 77} (2017) 47} [\href{https://arxiv.org/abs/1607.03412}{{\ttfamily
  1607.03412}}].

\bibitem{cta_sst1m}
{\scshape CTA-SST-1M project} collaboration,
  \href{https://doi.org/10.22323/1.358.0694}{\emph{PoS} {\bfseries ICRC2019}
  (2021) 694}.

\bibitem{tavernier_sst1m_2023}
{\scshape SST-1M} collaboration, {\emph{PoS} {\bfseries ICRC2023} (2023) 741}.

\bibitem{ruben_lopez_coto_2022_6344674}
R.~Lopez-Coto et~al.,  Mar., 2022.
\newblock 10.5281/zenodo.6344674.

\bibitem{karl_kosack_2021_5720333}
K.~Kosack et~al.,  Nov., 2021.
\newblock 10.5281/zenodo.5720333.

\bibitem{universe7100374}
C.~Nigro et~al., {\emph{Universe} {\bfseries 7} (2021) }.

\bibitem{axel_donath_2021_5721467}
A.~Donath et~al.,  Nov., 2021.
\newblock 10.5281/zenodo.5721467.

\bibitem{scikit-learn}
F.~Pedregosa, G.~Varoquaux, A.~Gramfort, V.~Michel, B.~Thirion, O.~Grisel
  et~al., {\emph{Journal of Machine Learning Research} {\bfseries 12} (2011)
  2825}.

\bibitem{1998cmcc.book.....H}
D.~{Heck} et~al., \emph{{CORSIKA: a Monte Carlo code to simulate extensive air
  showers.}} 1998.

\bibitem{BERNLOHR2008149}
K.~Bernlöhr,
  \href{https://doi.org/https://doi.org/10.1016/j.astropartphys.2008.07.009}{\emph{Astroparticle
  Physics} {\bfseries 30} (2008) 149}.

\bibitem{1985ICRC....3..445H}
A.~M. {Hillas},  in \emph{19th International Cosmic Ray Conference (ICRC19),
  Volume 3}, vol.~3 of \emph{International Cosmic Ray Conference}, p.~445,
  Aug., 1985.

\bibitem{LESSARD20011}
R.~Lessard, J.~Buckley, V.~Connaughton and S.~{Le Bohec},
  \href{https://doi.org/https://doi.org/10.1016/S0927-6505(00)00133-X}{\emph{Astroparticle
  Physics} {\bfseries 15} (2001) 1}.

\bibitem{Hofmann:1999rh}
W.~Hofmann, I.~Jung, A.~Konopelko, H.~Krawczynski, H.~Lampeitl and
  G.~P\"uhlhofer,
  \href{https://doi.org/10.1016/S0927-6505(99)00084-5}{\emph{Astropart. Phys.}
  {\bfseries 122} (1999) 135}
  [\href{https://arxiv.org/abs/astro-ph/9904234}{{\ttfamily
  astro-ph/9904234}}].

\bibitem{ALEKSIC201676}
J.~Aleksić et~al.,
  \href{https://doi.org/https://doi.org/10.1016/j.astropartphys.2015.02.005}{\emph{Astroparticle
  Physics} {\bfseries 72} (2016) 76}.

\bibitem{sst1m_mc_2019}
{\scshape CTA Consortium} collaboration,
  \href{https://doi.org/10.22323/1.358.0708}{\emph{PoS} {\bfseries ICRC2019}
  (2020) 708} [\href{https://arxiv.org/abs/1907.08061}{{\ttfamily
  1907.08061}}].

\bibitem{2020JInst..15P1010A}
C.~{Alispach}, J.~{Borkowski}, F.~R. {Cadoux}, N.~{De Angelis}, D.~{Della
  Volpe}, Y.~{Favre} et~al.,
  \href{https://doi.org/10.1088/1748-0221/15/11/P11010}{\emph{Journal of
  Instrumentation} {\bfseries 15} (2020) P11010}
  [\href{https://arxiv.org/abs/2008.04716}{{\ttfamily 2008.04716}}].

\bibitem{DeAngelis:2683289}
N.~De~Angelis, {\emph{Master thesis} (2019) }.

\bibitem{Nagai_2019}
A.~Nagai et~al.,
  \href{https://doi.org/10.1088/1748-0221/14/12/P12016}{\emph{Journal of
  Instrumentation} {\bfseries 14} (2019) P12016}.

\bibitem{ALEKSIC201530}
J.~Aleksić et~al.,
  \href{https://doi.org/https://doi.org/10.1016/j.jheap.2015.01.002}{\emph{Journal
  of High Energy Astrophysics} {\bfseries 5-6} (2015) 30}.

\bibitem{DeMitri:2021jog}
{\scshape DAMPE} collaboration,
  \href{https://doi.org/10.22323/1.358.0148}{\emph{PoS} {\bfseries ICRC2019}
  (2021) 148}.

\end{thebibliography}\endgroup

\end{document}